\documentclass[fleqn,usenatbib]{mnras}
\hypersetup{bookmarks=false}
\usepackage{newtxtext,newtxmath}

\usepackage[T1]{fontenc}
\usepackage{ae,aecompl}
\newcommand{\teff}{$T_\mathrm{eff}$}

\usepackage{graphicx}	
\usepackage{amsmath}	
\usepackage{amssymb}	

\title[]{Ultracool Dwarfs in deep extragalactic surveys using the Virtual Observatory: ALHAMBRA and COSMOS}

\author[Solano et al.]{
E.~Solano$^{1,2},$\thanks{E-mail: esm@cab.inta-csic.es}
M.~C.~G\'alvez-Ortiz$^{3}$,
E.~L.~Mart\'in$^{4,5,6}$,
I.~M.~G\'omez Mu\~noz$^{7}$,
\newauthor
C.~Rodrigo$^{1,2}$,
A.~J.~Burgasser$^{8}$,
N.~Lodieu$^{4,5}$,
V.~J.~S.~Béjar$^{4,5}$,
N.~Hu\'elamo$^{1}$,
\newauthor
M.~Morales-Calder\'on$^{1}$,
and
H.~Bouy$^{9}$
\\
$^{1}$Centro de Astrobiolog\'{\i}a (CSIC-INTA), ESAC Campus, Camino Bajo del Castillo s/n, E-28692, Villanueva de la Ca\~{n}ada, Madrid, Spain\\
$^{2}$Spanish Virtual Observatory\\
$^{3}$ Suffolk University, Madrid Campus, C/ Valle de la Vi\~{n}a 3, 28003,
Madrid, Spain \\
$^{4}$Instituto de Astrof\'isica de Canarias (IAC), Calle V\'ia L\'actea s/n, E-38200 La Laguna, Tenerife, Spain\\
$^{5}$Departamento de Astrof\'isica, Universidad de La Laguna (ULL), E-38206 La Laguna, Tenerife, Spain\\
$^{6}$Consejo Superior de Investigaciones Científicas, E-28006 Madrid, Spain\\
$^{7}$ Universidad Internacional de Valencia. Carrer del Pintor Sorolla 21, 46002, Valencia, Spain\\
$^{8}$ Center for Astrophysics and Space Science, University of California San Diego, 9500 Gilman Drive, La Jolla, CA 92092, USA \\
$^{9}$ Laboratoire d’astrophysique de Bordeaux, University of Bordeaux, CNRS, B18N, All\'ee Geoffroy Saint-Hilaire, 33615 Pessac, France
}

\begin{document}
\label{firstpage}
\pagerange{\pageref{firstpage}--\pageref{lastpage}}
\maketitle
\begin{abstract}
Ultracool dwarfs encompass a wide variety of compact stellar-like objects with spectra classified as late-M, L, T and Y. Most of them have been discovered using wide-field imaging surveys. The Virtual Observatory has proven to be of great utility to efficiently exploit these astronomical resources.
We aim to validate a Virtual Observatory methodology designed to discover and characterize ultracool dwarfs in deep extragalactic surveys like ALHAMBRA and COSMOS. Three complementary searches based on parallaxes, proper motions and colours, respectively were carried out. A total of 897 candidate ultracool dwarfs were found, with only 16 previously reported in SIMBAD. Most of the new UCDs reported here are likely late-M and L dwarfs because of the limitations imposed by the utilization of optical ($Gaia$ DR2 and r-band) data. 
We complement ALHAMBRA and COSMOS photometry with other catalogues in the optical and infrared using VOSA, a Virtual Observatory tool that estimates effective temperatures from the spectral energy distribution fitting to collections of theoretical models. 
The agreement between the number of UCDs found in the COSMOS field and theoretical estimations together with the low false negative rate (known UCDs not discovered in our search) validates the methodology proposed in this work, which will be used in the forthcoming wide and deep surveys provided by the Euclid space mission. Simulations of Euclid number counts for UCDs detectable in different photometric passbands are presented for a wide survey area of 15,000 square degrees, and the limitations of applicability of Euclid data to detect UCDs using the methods employed in this paper are discussed.
 
\end{abstract}




\section{Introduction}

Ultracool dwarfs (UCDs) are defined as objects with spectral types M7\,V or later and comprise very low mass stars, brown dwarfs (BDs) and planetary-mass objects. They represent about 15\% of the stellar
population in the solar neighbourhood \citep{Henry06} and their lifetimes make them the longest-lived not evolved objects of the universe. The M7\,V spectral type is just below the substellar boundary in the benchmark Pleiades cluster \citep[e.g.][]{Martin96, Rebolo96}, and it marks the  
beginning of a variety of changes that are seen to happen with decreasing effective temperature, in particular the appearance of dust clouds and the  marked increase of pressure-broadened neutral atomic lines of alkali elements \citep[e.g.][]{Jones97, Martin97, Helling08, Kirk99}. 

In addition to be key elements to properly understand the physical processes at the stellar/substellar boundary, UCDs are excellent candidates for exoplanet searches. As they are much fainter and smaller than solar-like stars, it is much easier to detect low-mass rocky planets orbiting around them \citep{Martin13}. TRAPPIST\,-1 \citep{Gillon17} has been the first successful example of this type of planetary system.  

Not many studies of stellar and substellar objects in extragalactic surveys can be found in the literature. The search of brown dwarfs in the UKIDSS DXS and UDS surveys by \cite{Lodieu09} and the pioneering work by \cite{Cuby99} using the NTT Deep Field and \cite{Liu02} using the IfA Deep Survey are some of the few exceptions to this. In fact, UCDs are normally treated as contaminants since, at high-redshift, galaxies and ultracool dwarfs show similar colours in the near infrared \citep{Wilkins14, Ryan16}. This, together with the photometric depth of the extragalactic surveys, makes them very interesting niches for the discovey of UCDs. In this paper we use the ALHAMBRA and COSMOS surveys to discover and characterize new UCDs. These two extragalactic surveys have a relatively large field of view and low extinction, which facilitates the discovery a significant number of UCDs and the determination of their physical parameters. Our main goal is to assess the effectiveness of a Virtual Observatory (VO)-based methodology that could be used in future, deeper and larger surveys like those planned with the Euclid\footnote{http://sci.esa.int/euclid/} space mission.

ALHAMBRA \citep[Advance Large Homogeneous Area Medium-Band Redshift Astronomical,][]{Moles08} is a deep photometric survey aimed at providing precise photometric redshifts and information on the Spectral Energy Distributions (SEDs) of thousands of galaxies and active galactic nuclei. ALHAMBRA was conducted in eight different regions of the sky, covering a total area of 4\,deg$^{2}$ and including overlapping sections of COSMOS, DEEP2, ELAIS, GOODS-N, SDSS and Groth survey fields. Observations were made at Calar Alto Observatory with the 3.5\,m telescope using LAICA (optical) and Omega-2000 (near-IR) instruments. These observations provided photometric information in 20 contiguous, equal-width, medium-band filters from 3500 to 9700 \AA \ plus the standard $J$, $H$ and $Ks$ near-infrared bands.The ALHAMBRA filter set and associated limiting magnitudes are given in Table 3 of \cite{Molino14}..

The COSMOS project pioneered the study of
galactic structures at intermediate to high redshifts as well as the evolution of the galaxy and AGN populations, thanks to the
unique combination of a large area and precise photometric redshifts. In this work we made use of the COSMOS2015\footnote{http://cdsarc.u-strasbg.fr/viz-bin/cat/J/ApJS/224/24} catalogue, which contains photometric information for 1.182.108 sources over the 2\,deg$^{2}$ COSMOS field. This version of the COSMOS catalogue was improved from previous versions by the addition of new $Y$$J$$H$$Ks$ images from the UltraVISTA-DR2 survey, $Y$-band images from Subaru/Hyper-Suprime-Cam, and infrared data from the Spitzer Large Area Survey with the Hyper-Suprime-Cam Spitzer legacy program.The COSMOS photometric bands and the averaged limiting magnitudes are given in Table 1 of \cite{Laigle16}. All ALHAMBRA and COSMOS magnitudes are given in the AB system.

 

This article is organized as follows: In Sect. \ref{sec:method}, we describe the methodology we have followed to separate stars from galaxies and to identify candidate UCDs among the stellar sources. In Sect. \ref{sec:known} we assess the robustness of our methodology by studying the fraction of known UCDs that have been recovered. Sect. \ref{sec:spec} presents the spectroscopic analysis made to confirm the UCD nature of some of our candidates. Sect. \ref{sec:sim} deals with simulations of UCD number counts in the Euclid surveys and how the lessons learnt from this work may be used to optimize the identification of UCDs in those fields, and, finally, Sect. \ref{sec:conclusions} contains the conclusions of this work.

\section{Methodology}
\label{sec:method}
\subsection{Sample selection}

The ALHAMBRA final catalogue\footnote{http://svo2.cab.inta-csic.es/vocats/alhambra/} provides astrometric, morphometric, photometric, redshift and quality information for 438\,661 sources (see the {\it Documentation} section for a detailed description of the catalogue contents). 

We first applied a morphometric filter to keep only stellar objects. For this, we made use of the {\it stell} and {\it Stellar$_{-}$Flag} parameters with the conditions stell$\geq$0.5 and Stellar$_{-}$Flag$\geq$0.5 \citep{Molino14}.  {\it stell} is the stellarity parameter implemented in Sextractor with values ranging from 0 (galaxy) to 1 (star). {\it Stellar$_{-}$Flag} represents a source-by-source statistical classification among stars and galaxies as described in \cite{Molino14}. Later, we used a photometric flag to remove saturated objects ({\it Satur$\_$Flag}=0), and, finally, we discarded  bona-fide extragalactic objects by keeping sources with values in the redshift  column ({\it zb$\_$1}) smaller than 0.5. This parameter can have associated large uncertainties so, in order not to loose any potential candidate UCD, we decided to adopt this very conservative criterion. This gave us a list of 54\,611 objects.


In the COSMOS catalogue we applied a morphometric filter using the keyword {\it TYPE} set to 1  and {\it zphot} equal to 0 or -99.9. This way we ended up with 37\,069 objects. 


\subsection{Hertzsprung--Russell diagram}

We cross-matched the 54\,611 ALHAMBRA objects with {\it Gaia} DR2 using a 3\,arcsec radius. We kept only counterparts with relative errors of less than 10 percent in $G$ and $G_{RP}$ and less than 20 per cent in parallax. The absolute {\it Gaia} magnitude in the $G$ band for individual stars was estimated using
\begin{equation}
  M_G = G + 5 \log{\varpi} +5,   
\end{equation}
where $\varpi$ is the parallax in arcseconds. In our case, the inverse of the parallax is a reliable distance estimator because we kept only sources with relative errors in parallax lower than 20 per cent \citep{Luri18}. 

\begin{figure}
        \includegraphics[width=\columnwidth]{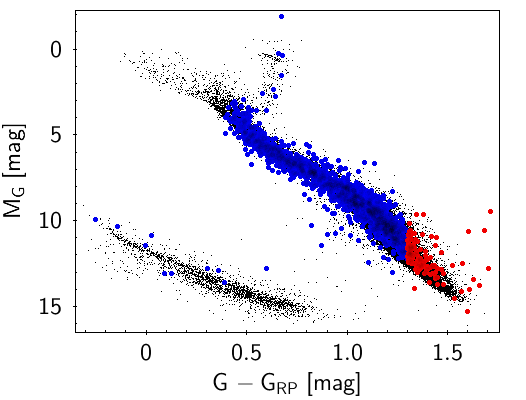}
    \caption{Colour-magnitude diagram built using {\it Gaia} DR2 sources with parallaxes larger than 10 mas and good photometry (small black dots). Large blue dots indicate the position of the 1\,548 ALHAMBRA stellar (dwarfs, giants and white dwarfs) sources. Candidate UCDs defined as objects with  $G$-$G_{RP}$ > 1.3 (119 objects) are overplotted in red. A similar distribution is obtained if COSMOS data are used.} 

        \label{fig:hrd}
\end{figure}

After the crossmatch  we ended up with 1\,548 stellar sources. A preliminary selection of UCDs was done using the updated version\footnote{https://www.pas.rochester.edu/$\sim$emamajek/} of Table 5 in \cite{Pecaut13} taking a conservative value of G-G$_{RP}$ > 1.3 (corresponding, according to this table, to M5\,V). This gave us 119 candidate UCDs. These candidates were plotted on top of a Hertzsprung--Russell diagram (HRD) built with all the {\it Gaia} DR2 objects at less than 100\,pc and good G,G$_{RP}$ photometry (Fig.\,\ref{fig:hrd}). This diagram is similar to that shown in fig. 6 of \cite{Babu18}.

\begin{figure}
\includegraphics[width=9cm]{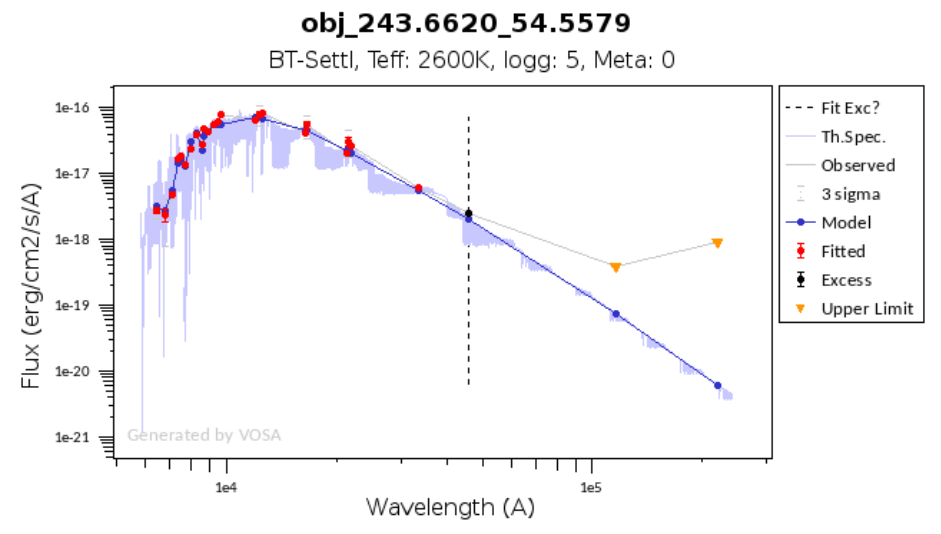}
    \caption{Example of an SED fitting as generated by VOSA. The blue spectrum represents the theoretical model that fits best, while red dots represent the observed photometry. The inverted yellow triangle indicates that the photometric value corresponds to an upper limit. 
}
        \label{fig:vosa}
\end{figure}

We made use of VOSA\footnote{http://svo2.cab.inta-csic.es/theory/vosa/} \citep{Bayo08} to estimate effective temperatures of these 119 objects. VOSA computes physical parameters by fitting the observational Spectral Energy Distribution (SED) to different collections of theoretical models. In our analysis we made use of the BT-Settl collection \citep{Allard12}. Gravity and metallicity were restricted to the ranges logg: 4.5 -- 6 and [M/H]: -0.5 -- +0.5, respectively. No extinction correction was used as low extinction was the first and basic criterion to select the ALHAMBRA fields \citep{Moles08}. An example of a VOSA SED fitting can be found in Fig.\,\ref{fig:vosa}.


To calibrate the effective temperatures obtained from the comparison with the BT-Settl models using VOSA with spectral types, we use the list of spectroscopically confirmed M dwarfs compiled by \cite{West11}. The mean effective temperature as a function of the spectral type as well as the standard deviations are given in Table~\ref{tab:Teff}. According to this table we adopt for UCDs a conservative value of \teff\, $\leq$\,2\,900\,K. Only sources with good SED fitting (vgfb<15) were kept. {\it vgfb} is a modified chi2, calculated by forcing $\Delta F_{obs}$ to be larger than 0.1$\times$F$_{obs}$, where $\Delta F_{obs}$ is the error in the observed flux (F$_{obs}$). This parameter is particularly useful when the photometric errors of any of the catalogues used to build the SED are underestimated. vgfb $<$ 15 is a reliable indicator of good fit. After applying these conditions we ended up with 30 sources.

Similarly, we cross-matched the 37\,069 objects in the COSMOS sample with {\it Gaia} DR2 using a 3\,arcsec radius. We found 207 fulfilling the conditions on parallaxes and photometry, out of which only one (COSMOS\,998096) has \teff\, $\leq$\,2\,900\,K. 

A mean value of $\sim$ 170\,pc was obtained for the 31 sources (30 ALHAMBRA + 1 COSMOS) with good parallaxes, with the closest and
farthest objects at 70 and 350\,pc, respectively.

\subsection{Proper motions}

Proper motions can be used to discriminate between galaxies and stars because extragalactic objects must have negligible values. For instance, M31, the nearest galaxy to the Milky Way, has a proper motion of just a few tenths of microarcseconds \citep{Marel19}. Taking the remaining 53\,063 (54\,611 - 1\,548) ALHAMBRA sources, we found 5\,102 sources having {\it Gaia} counterparts in a 3 arcsec radius, out of which 3\,774 are sources with non-zero proper motions, defining as such those sources fulfilling that, at least, one of the proper motion components is larger (in absolute value) than three times the associated error. Among them, 1\,733 sources also fulfilled the conditions of having relative errors of less than 10 per cent in $G$ and $G_{RP}$ and less than 20 per cent in both proper motion components.

\begin{figure}
        \includegraphics[width=\columnwidth]{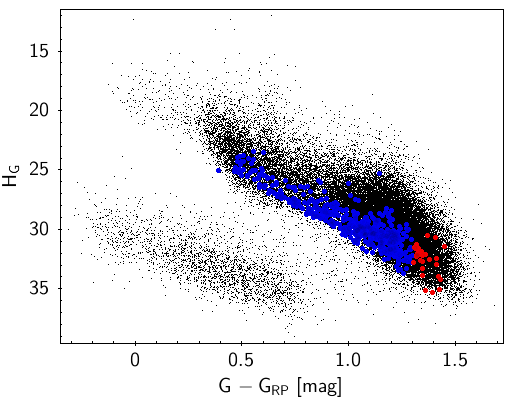}
\caption{Reduced proper motion -- colour diagram. In black we plot the objects used to build the Hertzsprung--Russell diagram in Fig.\,\ref{fig:hrd}. Blue circles represent the 462 ALHAMBRA sources that lie in the expected locus for dwarfs. Candidate UCDs defined as objects with  G-G$_{RP}$ > 1.3 (29 objects) are overplotted in red. A similar distribution is obtained if COSMOS data are used. }
        \label{fig:rpm}
\end{figure}

Of these 1\,733 sources, 462 lie in the expected locus for dwarfs in a reduced proper motion diagram defined as:
\begin{equation}
H_{G} = G + 5 \log{\mu} + 5
,\end{equation}
where $G$ is the {\it Gaia} apparent magnitude and $\mu$ is the total proper motion in mas yr$^{-1}$ (Fig.\,\ref{fig:rpm}). In this type of diagrams, proper motion is used as a proxy for distance assuming that nearby objects will have higher proper motions. Just as in a Hertzsprung--Russell diagram, reduced proper motion diagrams have been shown to be excellent tools for segregating the various stellar populations 
\citep{Dhital10, Jimenez11, Lodieu12, Zhang13, Lodieu17, Smart17}. 

Among the 462 sources, 29 also fulfill the $G$-$G_{RP}$ > 1.3 condition.  We applied VOSA to this sample, keeping only those with \teff\,$\leq$\,2\,900\,K. We ended up with just one object (ALHAMBRA 81481207114). 

Regarding COSMOS, 678 sources satisfy the conditions on magnitudes and proper motion components, out of which 59 also fulfill the  $G$-$G_{RP}$ > 1.3 condition. After applying VOSA we kept only one source with \teff\, $\leq$\,2\,900\,K (COSMOS 690108).

\begin{table}
	\centering
	\caption{Effective temperatures calculated using BT-Settl models as a function of the spectral type of the objects given in \citep{West11}}
	\label{tab:Teff}
	\begin{tabular}{cccc} 
		\hline
		Spectral & Number of & Effective  & Standard\\
        Type     & objects  & temperature (K)  & deviation (K)  \\
		\hline
		M5 & 3901 & 3213 & 206\\
		M6 & 5645 & 3057 & 106\\
		M7 & 5824 & 2959 & 152\\
        M8 & 1682 & 2715 & 157\\
        M9 & 891  & 2596 & 162 \\
		\hline
	\end{tabular}
\end{table}

\subsection{Photometry}


After applying the first two criteria (location in the H--R diagram and reduced proper motion), only 3\,281 sources (1\,548 +  1\,733, respectively) have been classified as stars in the ALHAMBRA survey. To discriminate between stars and galaxies in the remaining list of 51\,330 (54\,611 - 3\,281) sources, we used the ALHAMBRA photometric information. We assigned a "true galaxy" status to all objects with {\it Stellar$\_$Flag} and {\it stell} in the range 0.0 -- 0.1, and a "true star" status to all objects with these flags in the range 0.9 -- 1. After testing different combinations, we decided to use the $J-Ks$ vs $F644W-F923W$ colour-colour diagram to discriminate between the two types of sources. This is similar to the criterion used in fig. 9 of  \cite{Strait15}, but adapted to ALHAMBRA photometry. 

\begin{figure*}
        \includegraphics[width=\columnwidth]{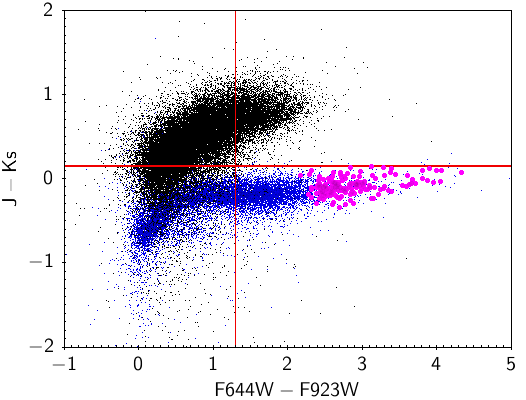}
        \includegraphics[width=\columnwidth]{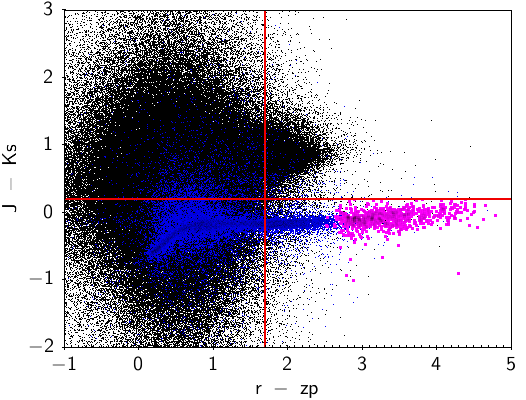}
    \caption{ALHAMBRA (left) and COSMOS (right) colour-colour diagrams. Black and blue dots represent galaxies and stars, respectively. The vertical and horizontal red lines mark the boundaries of the stellar and galactic locus. Pink bullets represent the 164 (30+1+133) ALHAMBRA and the 733 (1+1+731) COSMOS candidate UCDs. 
}
        \label{fig:phot}
\end{figure*}

Fig.\,\ref{fig:phot} shows the ALHAMBRA and COSMOS "true star" and "true galaxy" objects in the colour-colour diagram. Following  \cite{Strait15}, the region in the colour-colour diagram fulfilling the following conditions:
\begin{equation}
\begin{gathered}
F644W-F923W > 1.3 \\
 J-Ks <0.15
\end{gathered}
\end{equation}

is defined as the "ALHAMBRA true star locus free of contamination (< 1\,per cent) by extragalactic object".

Likewise, the region above $J-Ks$ > 0.15 is defined as the extragalactic locus free of stellar contamination (<1\,per cent). The remaining region

\begin{equation}
\begin{gathered}
F644W-F923W \leq 1.3 \\
 J-Ks \leq 0.15
\end{gathered}
\end{equation}

is occupied both by galaxies (42\%) and stars (58\%) and an efficient separation cannot be applied. 

Using this criteria with the remaining 51\,330 sources, we were able to classify 15\,353 sources as galaxies, 1\,402 as stars while 14\,846 sources lie in the overlapping region between stars and galaxies and were not classified. The remaining 19\,729 sources had bad photometry in any of the four filters and were not classified either. 

According to the position in the colour-colour diagram of the UCDs found in the two previous sections, we define as candidate UCD a stellar source with $F644W-F923W$ > 2.15. A total of 272 sources passed this criterion. After applying VOSA, we kept 133 with Teff $\leq$ 2\,900\,K. 


Similarly, we selected all COSMOS sources with good photometry in the $r$, $zp$ (Subaru) and $J$, $Ks$ (ULTRAVISTA) bands and separate stars from galaxies in a $(J-Ks)$ vs $(r-zp)$ colour-colour diagram. We used the flags {\it TYPE} to separate stars from galaxies. The stellar locus was defined by the following region (Fig.\,\ref{fig:phot}).

\begin{equation}
\begin{gathered}
r-zp \geq 1.7 \\
 J-Ks \leq 0.2
\end{gathered}
\end{equation}

We found 1\,220 sources fulfilling these criteria, out of which 731 have effective temperatures $\leq$ 2\,900\,K. 
Adding the candidate UCDs found by the three searches (HRD, proper motion and colours) a total of 897 (164 ALHAMBRA + 733 COSMOS) sources is obtained. 

\begin{figure}
    \includegraphics[width=\columnwidth]{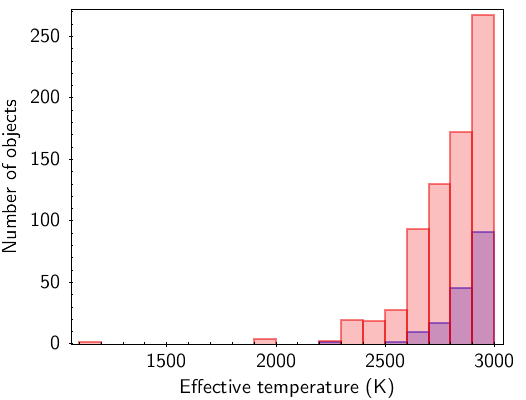}
    \caption{\teff$ $ distribution of the candidate UCDs identified in COSMOS (red) and ALHAMBRA (blue).}
    \label{fig:teffcand}
\end{figure}

Fig.\,\ref{fig:teffcand} shows the distribution of effective temperatures of our 897 candidate UCDs. Most of them lie in the 2\,600 -- 2\,900\,K range while the coolest object (COSMOS 752243 / CFBDS J100113+022622, a T5 brown dwarf) shows a \teff\, = 1\,100\,K. The full list of 897 candidates is available online at the SVO archive of ultracool dwarfs identified in ALHAMBRA and COSMOS (see Appendix A).

\begin{figure}
        \includegraphics[width=\columnwidth]{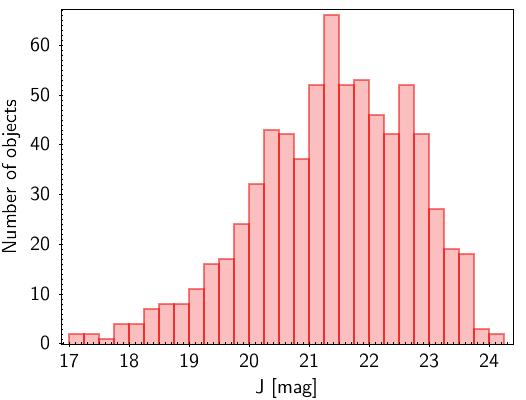}
    \caption{Distribution of the COSMOS candidate UCDs magnitudes. UltraVISTA DR2 $J$-band (AB) magnitude in a 3" aperture have been considered. 
}
        \label{fig:jhist}
\end{figure}

Different studies have been devoted to estimate the surface density of cool stars and brown dwarfs in deep extragalactic surveys, mainly to account for the degree of contamination by these objects due to their similar near-infrared colors to high redshift galaxies \citep[e.g.,][]{Wilkins14}. In particular, \cite{Ryan16} calculate the surface density of ultracool dwarfs in representative fields of the James Webb Space Telescope, including COSMOS. Using their tables\,4--7  and fig.\,6, and assuming $J$ =23.5 as COSMOS limiting magnitude (Fig.\,\ref{fig:jhist}), we compared our list of candidate UCDs to their estimations finding a reasonable good agreement (Table~\ref{tab:cosmosrr}).

\section{Known ultracool dwarfs in the ALHAMBRA and COSMOS fields}
\label{sec:known}

In this section we assess the fraction of known UCDs that have been recovered using our methodology. In particular we looked for UCDs in SIMBAD \citep{Wenger00}.

Using the SIMBAD TAP service\footnote{\tt http://simbad.u-strasbg.fr:80/simbad/sim-tap}, we chose objects having spectral types later than M7V or labelled as {\it brown dwarfs}. A total of 13\,329 objects fulfilled these conditions. 

To know how many of them lie in the region of the sky covered by ALHAMBRA and COSMOS, we took advantage of Aladin and the Multi-Object Coverage\footnote{\tt http://ivoa.net/documents/MOC/index.html} VO standard. Regarding ALHAMBRA, a total of 193 SIMBAD UCDs lie in its field of view, but only 10 were really included in the ALHAMBRA catalogue. For COSMOS, there were 15 sources in the field of view, out of which 12 were in \cite{Laigle16}.

The efficiency of our search was estimated using the false negative rate (number of known UCDs in SIMBAD that were not rediscovered in our search). For ALHAMBRA, nine of the objects were recovered by us. The remaining object (ALHAMBRA 81441106044) escaped from our search because of its low {\it stell} value (0.1). For COSMOS, seven objects were recovered while the remaining five escaped from our search due to different reasons: {\it TYPE} value different from 1 (2 sources), lack of $Ks$ photometry (1 source) and $(J-Ks)$ colour slightly higher than 0.15 (2 sources). 


\section{Spectroscopic follow-up}
\label{sec:spec}

\begin{figure*}
\centering
\includegraphics[width=\textwidth]{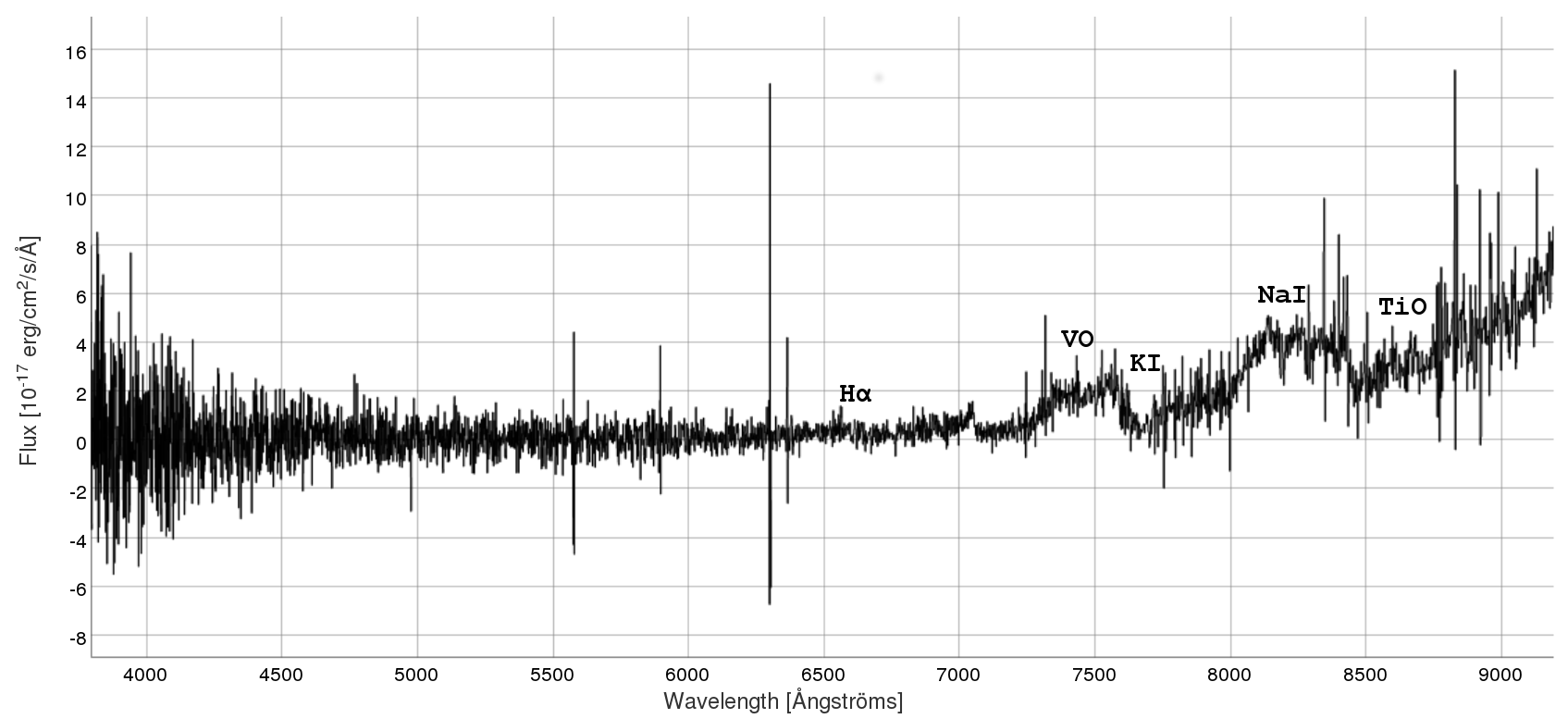}
\caption{SDSS spectra of COSMOS\,247323 (M9V). Together with the rest of SDSS spectra (12), it is available from the SVO archive (see Appendix).}
\label{fig:sdssspec}
\end{figure*}

Another way to confirm the validity of our methodology to find new UCDs is to look for available spectroscopic information in public archives. We found spectra for 11 ALHAMBRA candidate UCDs in the SDSS DR16 archive\footnote{https://dr16.sdss.org} using a 10\,arcsec search radius. SDSS spectra cover the optical range ($\sim$3800 -- 9200 \AA)  with a resolving power $\lambda/(\Delta\lambda)$ = 1800. Spectra were automatically reduced and assigned a spectral type by the SDSS pipeline software following \cite{Bolton12}.  In all cases the sources were classified as {\it "stars"} with spectral types M5 (4), M6 (5), and M7 (2) . We can see how the conservative value of \teff$ $ adopted to select UCDs has a clear impact on the non-negligible number of contaminants (i.e. objects with spectral types earlier than M7\,V). Two of the M6 objects 
(ALHAMBRA\,81461309369 and ALHAMBRA\,81473108856) were classified in SDSS as M6\,III. However, the position of the first source in the HRD clearly indicates that this object is a dwarf. The second target was photometrically selected and we cannot conclude about its dwarf/giant nature.
Regarding COSMOS we found spectroscopic information for two objects, COSMOS\,247323 and COSMOS\,601887, classified as M9\,V and M6\,V, respectively (Fig.\,\ref{fig:sdssspec}). The list of objects with spectral type in SDSS are given in Table~\ref{tab:sptyp}.  One more object (ALHAMBRA\,81422407651, classified as a M3 star) was found in LAMOST DR5.


Additionally, we got service mode time with LIRIS\footnote{https://www.ing.iac.es/Astronomy/instruments/liris/}\, \citep{Manchado98} for two ALHAMBRA objects at the 4.2\,m William Herschel Telescope (WHT) in La Palma Observatory. The spectra of ALHAMBRA\,81473108856 and ALHAMBRA\,81473110590 were obtained on 25--27th June 2018 with a clear sky and a seeing of 0.9\,arcsec for 81473108856 and 0.7\,arcsec for 81473110590, respectively. The observations were reduced in the standard way (sky subtraction, flat-field division, extraction of the spectra and wavelength calibration) with IRAF\footnote{IRAF is distributed by the National Optical Observatory, which is operated by the Association of Universities for Research in Astronomy, Inc., under contract with the National Science Foundation.}. The spectra have a wavelength coverage of $\sim$8890 -- 23600 \AA \ with a resolving power $\lambda/(\Delta\lambda)$$ \approx$ 2600.  

We used a service of the Spanish Virtual Observatory\footnote{http://svo2.cab.inta-csic.es/theory/newov2/} to gather observational and theoretical templates covering the M6 -- L8 spectral range. These templates were used to assess the UCD nature of the candidates . In particular, we used the NIRSPEC\ library \citep{McLean03, McLean07}. The templates were first convolved to match the LIRIS spectral resolution and the comparison was done by visual inspection. 

\begin{figure}
\centering
\includegraphics[width=\columnwidth]{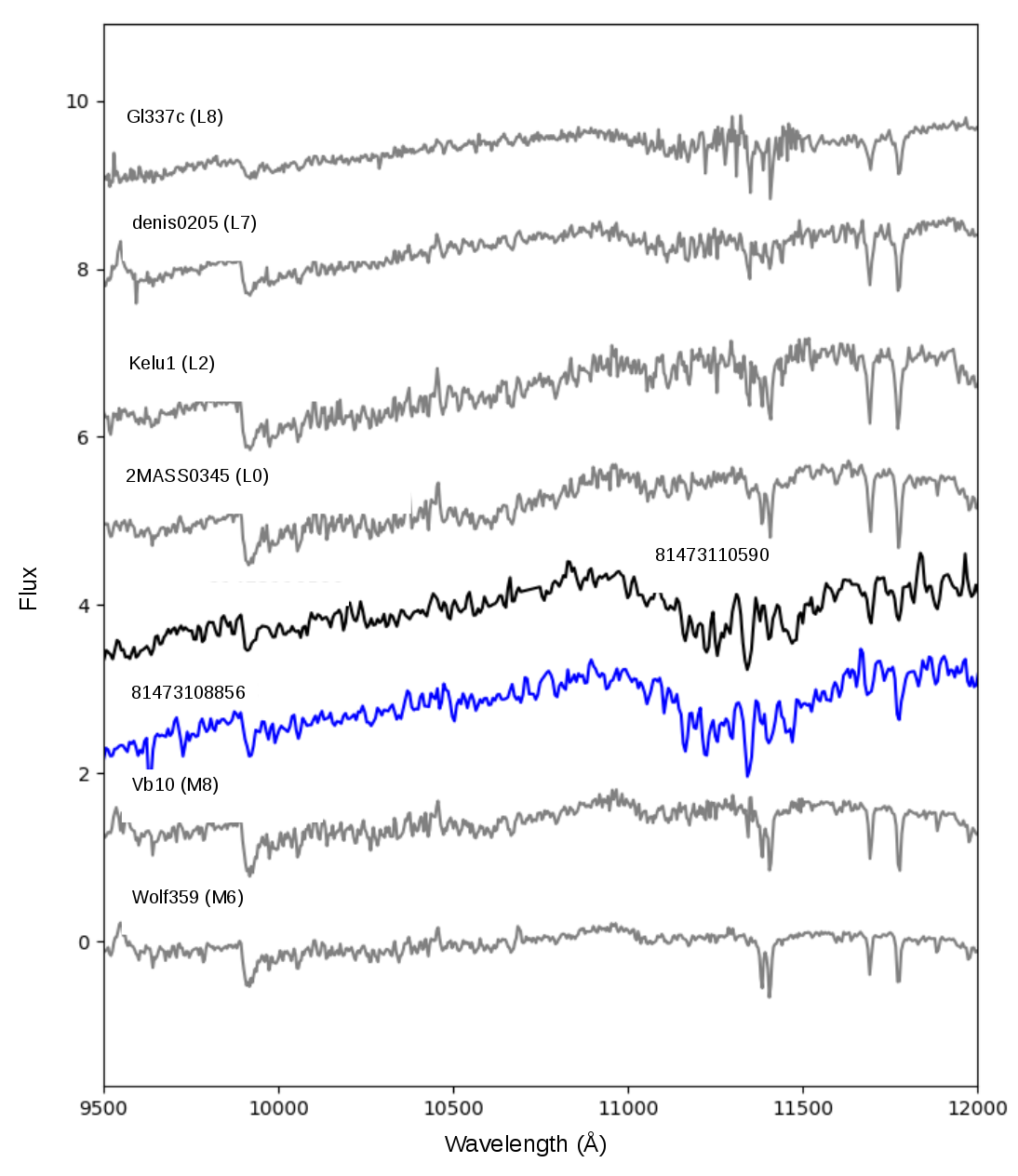}
\caption{Comparison of the LIRIS spectra of two of our ALHAMBRA UCD candidates with template spectra from the NIRSPEC library.}
\label{fig:lirisespeccomp1}
\end{figure}

For visual comparison, we plot the LIRIS spectra together with those of the NIRSPEC library in the $Y$ and $J$-bands, which is where our targets have the best signal to noise ratio ranging from $\sim$30 to $\sim$40 (Fig.~\ref{fig:lirisespeccomp1}). The presence of the $K$~{\sc i} doublet at 1.168\,$\mu$m and 1.177\,$\mu$m clearly indicates that our targets are UCDs. 
We defer to a later paper the detailed analysis of these spectra and additional follow-up spectra of UCDs that we plan to obtain for preparation of the Euclid mission.

\begin{table*}
\caption[]{Candidate UCDs with SDSS spectral types. The column {\it method} indicates whether the candidate has been selected using their position in the HR diagram (HRD) or their colours (PHOT). }
\label{tab:sptyp}
\begin{flushleft}
\begin{center}
\begin{tabular}{|r|r|l|l|r|r|r|r|l|r|c|}
\noalign{\smallskip}
\hline
 \multicolumn{1}{|c|}{RA} &
  \multicolumn{1}{c|}{DEC} &
  \multicolumn{1}{c|}{Survey} &
  \multicolumn{1}{c|}{Source\_id} &
  \multicolumn{1}{c|}{$\varpi$} &
  \multicolumn{1}{c|}{$\mu_{\alpha}\cos{\delta}$} &
  \multicolumn{1}{c|}{$\mu_{\delta}$} &
  \multicolumn{1}{c|}{Gmag} &
  \multicolumn{1}{c|}{Method} &
  \multicolumn{1}{c|}{\teff} &
  \multicolumn{1}{c|}{Sp. Type} \\
       (ICRS, deg) & (ICRS, deg) & & & [mas]& [mas/yr] & [mas/yr] &[mag] & & \\
\noalign{\smallskip}
\hline
  37.0651 & 0.6742 & ALHAMBRA & 81421305057 &  &  &  &  & PHOT & 2600 & M7\\
  37.3364 & 0.5783 & ALHAMBRA & 81422407651 & 5.96 & 21.60 & -8.08 & 18.80 & HRD & 2900 & M5\\
  37.3496 & 0.6364 & ALHAMBRA & 81422405585 & 7.43 & 84.99 & -16.34 & 18.46 & HRD & 2900 & M5\\
  139.4752 & 45.9878 & ALHAMBRA & 81431401913 & 14.22 & -36.03 & -20.49 & 18.75 & HRD & 2700 & M7\\
  149.6099 & 2.2107 & COSMOS & 601887 &  &  &  & 19.87 & PHOT & 2900 & M6\\
  149.9955 & 1.65567 & COSMOS & 247323 &  &  &  & 20.61 & PHOT & 2500 & M9\\
  150.3242 & 2.3860 & ALHAMBRA & 81441108312 &  &  &  & 19.78 & PHOT & 2800 & M6\\
  188.4841 & 61.7146 & ALHAMBRA & 81451310113 & 5.52 & -8.67 & -34.86 & 18.98 & HRD & 2900 & M5\\
  213.3649 & 52.8089 & ALHAMBRA & 81462203395 & 12.30 & 5.87 & -45.38 & 17.79 & HRD & 2800 & M6\\
  214.0596 & 52.1862 & ALHAMBRA & 81461309369 & 7.57 & 25.29 & -50.40 & 18.90 & HRD & 2900 & M6III\\
  214.6764 & 52.7097 & ALHAMBRA & 81461107375 & 2.86 & -6.64 & 8.33 & 20.15 & HRD & 2900 & M5\\
  214.9342 & 52.3238 & ALHAMBRA & 81461402862 & 7.08 & -33.87 & 19.14 & 18.33 & HRD & 2800 & M6\\
  243.7100 & 54.5900 & ALHAMBRA & 81473108856 &  &  &  & 21.14 & PHOT & 2600 & M6III\\ 
\noalign{\smallskip}
\hline
\noalign{\smallskip}
\end{tabular}
\end{center}
\end{flushleft}
\end{table*}

\begin{table}
\caption[]{COSMOS UCDs number counts. \cite{Ryan16} estimations have been computed considering both the thin and thick disc components and an sky area of 2 deg$^{2}$.}
\label{tab:cosmosrr}
\begin{flushleft}
\begin{center}
\begin{tabular}{lcc}
\noalign{\smallskip}
\hline
\noalign{\smallskip}
Spectral type & \cite{Ryan16}  & This work\\
\hline
 M8--M9 & 100 & 120 \\
 L0--L9 & 72 & 43 \\
 T0--T5 & 1--2 & 1 \\
\noalign{\smallskip}
\hline
\noalign{\smallskip}
\end{tabular}
\end{center}
\end{flushleft}
\end{table}
\section {Simulations of ultracool dwarf detections in the EUCLID wide survey}
\label{sec:sim}

This decade the Euclid Space Mission will cover 15,000 square degrees of the extragalactic sky with just one single epoch (the Euclid wide survey) and 40 square degrees in three selected regions that will be observed repeatedly during the lifetime of the mission (six years).

The COSMOS field is likely to be selected as a calibration field for Euclid, and hence the analysis of UCDs considered in this work could be used as reference. 

The requirements for Euclid were stated in the Red Book \citep{Lau11}. In this work an updated version of the Euclid filter transmission passbands\footnote{http://svo2.cab.inta-csic.es/theory/fps/index.php?\&gname=Euclid} has been adopted as follows: photometric sensitivity 24.5 mag. (AB) in VIS passband (0.55--0.9 $\mu$m), and 24 mag. (AB) in three NIR passbands, Y (0.945--1.231$\mu$m), J (1.159--1.587$\mu$m), and H (1.510--2.000$\mu$m), and slit-less spectroscopic sensitivity of 21 mag. (AB) at spectral resolution of 250 over the wavelength range 1.25 -- 1.85$\mu$m. The corresponding sensitivities for the deep survey are 2 mag. deeper. 

The simulation parameters are the following: Log-normal mass function with $\alpha$ parameter $=$0.5 \citep{Chabrier05}, star-formation rate from \cite{Aumer09}, evolutionary models from \cite{Burrows93}, spectral type versus absolute magnitudes and Teff from \cite{Pecaut13}, disk scale heights from 250 to 450 pc \citep{Ryan16}, although a value around 450 pc appears more likely from recent results by \cite{Sorahana19} and \cite{Carnero19}, and constant galactic latitude at 45 degrees for all objects and total survey area of 15,000 square degrees. 

\begin{figure}
    \includegraphics[width=\columnwidth]{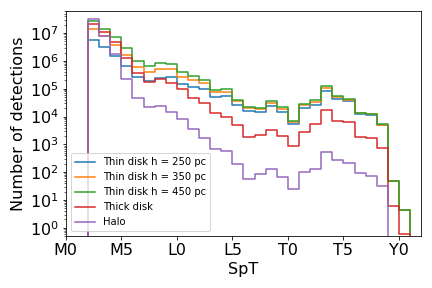}
    \caption{Simulated number counts of UCDs detected by the Euclid wide survey (15,000 square degrees) in the NISP J-band for a constant galactic latitude of 45 degrees for all objects.}
    \label{fig:wideJ}
\end{figure}

\begin{figure}
    \includegraphics[width=\columnwidth]{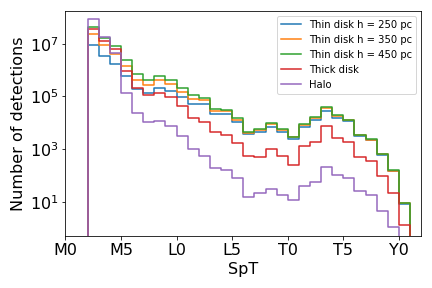}
    \caption{Same as Figure 9, but for the NISP Y-band.}
    \label{fig:wideY}
\end{figure}

\begin{figure}
    \includegraphics[width=\columnwidth]{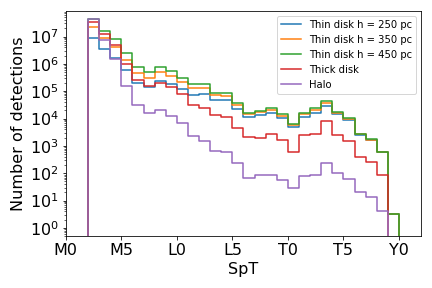}
    \caption{Same as Figure 9, but for the NISP H-band.}
    \label{fig:wideH}
\end{figure}

The most sensitive filter to UCDs in the Euclid wide survey will be the J-band as shown in Fig.~\ref{fig:wideJ}. About two million L dwarfs, one million T dwarfs and a  handful of Y dwarfs are expected to be detected in the thin disk for a scale height of 450 pc (green line). The corresponding numbers are about a factor of 2 lower for a scale height of 250 pc (blue line), about a factor of 10 lower for the thick disk population (red line), and about a factor of 500 lower for the halo (purple line).   For comparison, simulations of UCD number counts in the Y-band and H-band are shown in Fig.~\ref{fig:wideY} and 
Fig.~\ref{fig:wideH}, respectively. A very large number of UCDs (of order of 1 million objects) are expected to be detected in the three NISP passbands. This is important because, as shown in this work and in other works such as \cite{Deacon18, Holwerda18}, it is not enough to detect a UCD in one or two filters to identify it as such. Parallaxes, proper motions and/or multi-colors are needed. Ideally all of them are required for secure identification. 

\begin{figure}
    \includegraphics[width=\columnwidth]{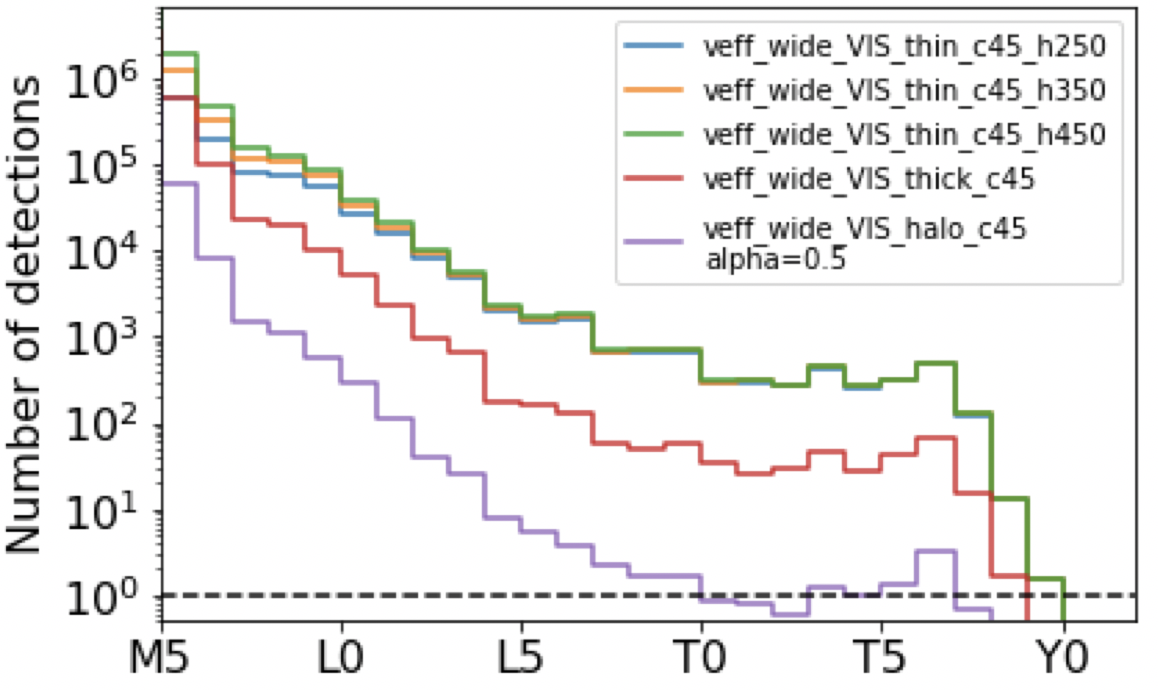}
    \caption{Same as Figure 9, but for the Euclid VIS-band.}
    \label{fig:wideVIS}
\end{figure}

The VIS instrument will provide higher spatial resolution than NISP, which will allow to resolve hundreds of UCD binaries. Simulations of UCD number counts in the VIS filter are shown in Fig.~\ref{fig:wideVIS}. The numbers of UCDs expected to be detected in the VIS-band are more than a factor of 100 less than in the J-band, particularly for late-T dwarfs. 

In the Euclid photometric catalogs, the availability of colors for UCDs is going to be provided mainly by the NISP instrument. The color-method will be based on Y-J versus J-H color color diagrams. The UCDs identified in this work will be useful to calibrate those diagrams with known UCDs. It will also be useful to cross identify Euclid point sources detected in the J band with the NEOWISE catalog in the W1 and W2 bands to find T and Y dwarfs. 

\begin{figure}
    \includegraphics[width=\columnwidth]{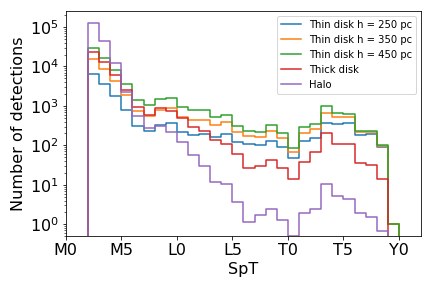}
    \caption{Same as Figure 9 but for the Euclid deep surveys (40 square degrees) in the NISP J-band.}
    \label{fig:deepJ}
\end{figure}

The proper motion and parallax methods of UCD identification can be applied in the Euclid deep fields and in the calibration fields. In Fig.~\ref{fig:deepJ} the predicted number counts of UCDs in the total area expected to be covered by the deep fields are presented. The number of T dwarfs that could be identified in the Euclid deep fields by proper motions in the J-band may be of order of a few thousands. These objects will be selected independently from their colors, and hence they will provide useful feedback to the color methods used to select UCDs in the wide field survey.

\section{Conclusions}
\label{sec:conclusions}
Using a Virtual Observatory methodology we have built a catalogue of 897 candidate UCDs found in the ALHAMBRA and COSMOS extragalactic surveys. Sixteen of them were already known in SIMBAD. Our primary goal in this paper was not to carry out a detailed analysis on the properties of the found candidates but to define and assess a search methodology to be used for deeper and larger surveys like Euclid, whose excellent sensitivity makes it an ideal resource for the discovery of new UCDs, including brown dwarfs with very low effective temperatures.

The use of different approaches based on parallaxes, proper motions and colours tends to minimize the drawbacks and biases associated to the search of ultracool objects: photometric-only selected samples may leave out peculiar UCDs not following the canonical trend in colour-colour diagrams and they can also be affected by extragalactic contamination. Proper motion searches may ignore objects with small values of projected velocity in the plane of the sky. Regarding parallax-based searches, they will be limited to the brightest objects with parallax values from {\it Gaia}.

With the help of the VOSA we estimated effective temperatures for our candidate UCDs. They range from 1\,100\,K to 2\,900\,K with the great majority of the objects in the 2\,600 -- 2\,900\,K range. We also compared the number of candidate UCDs found in the COSMOS field with theoretical estimations \citep{Ryan16}, finding a good agreement. 
The high success ratio recovering known UCDs demonstrates the robustness of our procedure, and the consistency with predicted counts of UCDs in the COSMOS fields indicates that the VO-based procedures described in this paper are reliable to efficiently mine forthcoming wide and deep surveys for new UCDs. We found that our choice of 2\,900\,K as an upper effective temperature limit in our selection is not restrictive enough to prevent somewhat earlier M dwarfs (M4--M6 V) from contaminating the sample of UCD candidates and will be taken into account in subsequent studies.  

Last but not least, we present simulated number counts of UCDs detected in the Euclid wide and deep fields, and we discuss the applicability of the methods of UCD detection used in this work.

\section*{Acknowledgements}
This research has made use of the Aladin sky atlas developed at CDS, Strasbourg Observatory, France. This publication makes use of VOSA, developed under the Spanish Virtual Observatory project supported from the Spanish MICIU through grant AyA2017-84089. This research has been partly funded by the Spanish State Research Agency (AEI) Project MDM-2017-0737 at Centro de Astrobiología (CSIC-INTA), Unidad de Excelencia María de Maeztu. EM was supported by grant AYA2015-69350-C3-1-P. NL was supported by grant AYA2015-69350-C3-2-P. This  work  has  made  use  of  data  from the  European Space Agency (ESA) mission {\it Gaia}\footnote{\tt https://www.cosmos.esa.int/gaia}, processed by the {\it Gaia} Data Processing and Analysis  Consortium  (DPAC)\footnote{\tt https://www.cosmos.esa.int/web/gaia/dpac/consortium}. We also extensively made use of TOPCAT \citep{Taylor05} and STILTS \citep{Taylor06} as well as the Vizier and SIMBAD services, both operated at CDS, Strasbourg, France.

\section{Data Availability Statement}

The data underlying this article are available at http://svo2.cab.inta-csic.es/vocats/alhambra\_cosmos/

\bibliographystyle{mnras}
\bibliography{ref_cosmos.bib}

\appendix

\section{Virtual Observatory compliant, online catalogue}

In order to help the astronomical community on using our
catalogue of candidate UCDs, we developed an archive system  that  can  be  accessed  from  a  webpage\footnote{\tt http://svo2.cab.inta-csic.es/vocats/alhambra\_cosmos/} or  through  a
Virtual Observatory ConeSearch

\begin{figure*}
\includegraphics[width=\textwidth]{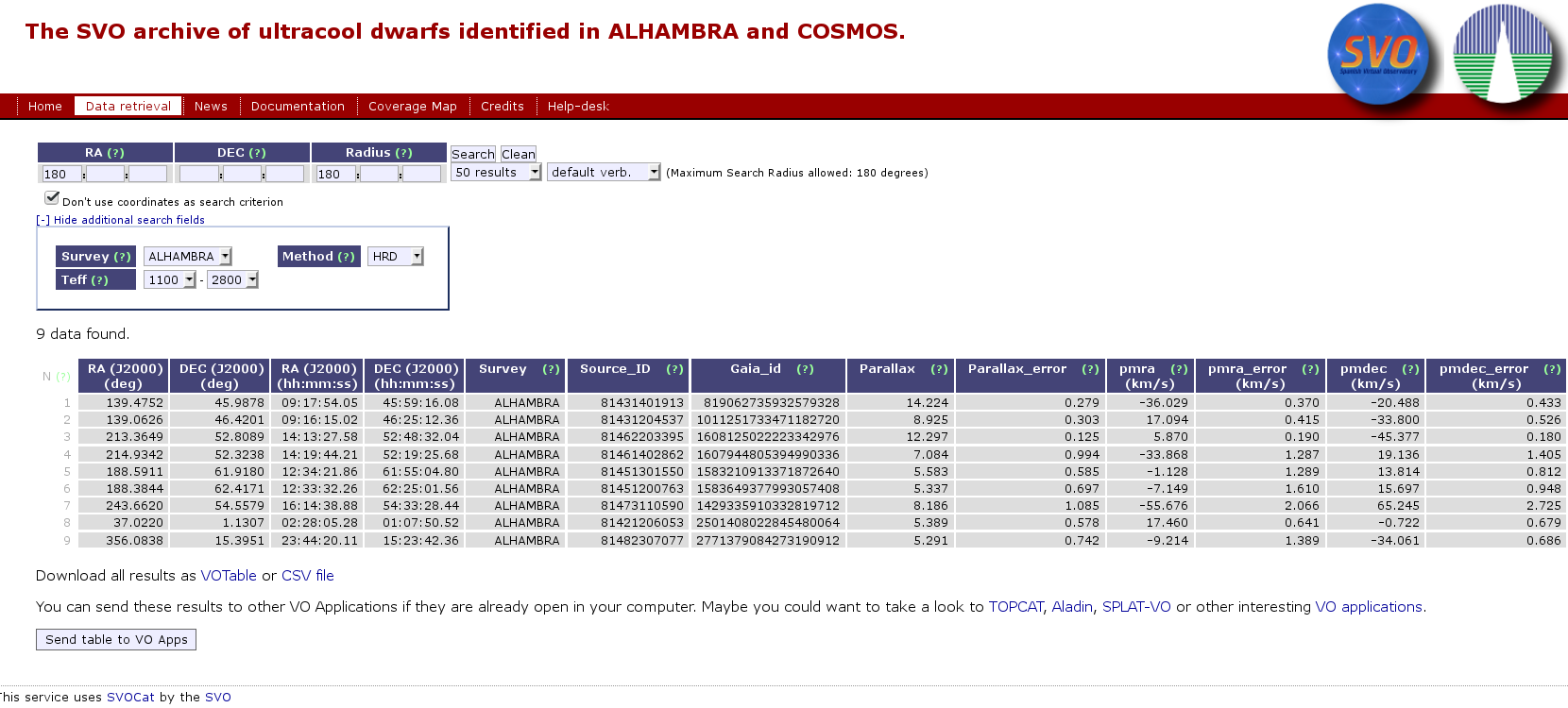}
\caption{Result from a query in the SVO archive of ultracool dwarfs identified in ALHAMBRA and COSMOS.
}
        \label{fig:archive}
\end{figure*}

The  archive  system  implements  a  very  simple  search
interface that allows queries by coordinates and radius as
well as by other parameters of interest. The user can also select the maximum number of sources (with values from 10 to
unlimited) and the number of columns to return (minimum,
default, or maximum verbosity).
The result of the query is a HTML table with all the
sources found in the archive fulfilling the search criteria (Fig.\,\ref{fig:archive}). The
result can also be downloaded as a VOTable or a CSV  file.
Detailed information on the output  fields can be obtained
placing the mouse over the question mark located close
to the name of the column. The archive also implements the
SAMP\footnote{\tt http://www.ivoa.net/documents/SAMP}
(Simple  Application  Messaging)  Virtual  Observatory  protocol.  SAMP  allows  Virtual  Observatory  applications  to  communicate  with  each  other  in  a  seamless  and
transparent manner for the user. This way, the results of a
query  can  be  easily  transferred  to  other  VO  applications,
such as, for instance, Topcat.

\

\bsp	
\label{lastpage}
\end{document}